\documentclass[aps,twocolumn,prl,showpacs,nofootinbib]{revtex4}

\usepackage{amsmath}
\usepackage{graphicx}
\usepackage{amssymb}

\newcommand{\vect}[1]{\boldsymbol{#1}}
\newcommand{\umode}[2]{u_{#1,#2}}
\newcommand{\umodeflat}[2]{u^\flat_{#1,#2}}
\newcommand{\green}[1]{G_\xi\negthinspace\left(#1\right)}
\newcommand{\Ylm}[2]{{Y_{#1}^{#2}}}
\newcommand{\besselJ}[1]{J{_{#1}}}
\newcommand{\laplace}[1]{\mathcal{L}\negthinspace\left\{#1\right\}}

\newcommand{\uc}{\mathrm{c}}
\newcommand{\ud}{\mathrm{d}}
\newcommand{\ue}{\mathrm{e}}
\newcommand{\uh}{\mathrm{h}}
\newcommand{\ub}{\mathrm{b}}
\newcommand{\ug}{\mathrm{g}}

\newcommand{\calD}{\mathcal{D}}

\newcommand{\Mpbulk}{M_*}
\newcommand{\Mp}{M_{\mathrm{Pl}}}

\newcommand{\ntot}{n}
\newcommand{\nextra}{\ntot_{\uc}}

\newcommand{\brane}[1]{\bar{#1}}
\newcommand{\Rbrane}{\brane{R}}
\newcommand{\gbrane}{\brane{g}}

\newcommand{\dilaton}{\psi}
\newcommand{\dilapot}{U}

\newcommand{\vev}{v}
\newcommand{\higgsvect}{\vect{\Phi}}
\newcommand{\higgs}{\phi}
\newcommand{\gauge}{C}

\newcommand{\md}{m_\ud}
\newcommand{\mh}{m_\uh}
\newcommand{\mb}{m_\ub}
\newcommand{\mg}{m_\ug}

\newcommand{\const}{C}

\begin{document}

\title{Massive gravitons trapped inside a hypermonopole}

\author{Antonio De Felice} \email{antonio.defelice@uclouvain.be}
\affiliation{Theoretical and Mathematical Physics Group, Centre for
  Particle Physics and Phenomenology, Louvain University, 2 Chemin du
  Cyclotron, 1348 Louvain-la-Neuve (Belgium)}

\author{Christophe Ringeval}
\email{christophe.ringeval@uclouvain.be}
\affiliation{Theoretical and Mathematical Physics Group, Centre for
  Particle Physics and Phenomenology, Louvain University, 2 Chemin du
  Cyclotron, 1348 Louvain-la-Neuve (Belgium)}

\date{\today}

\begin{abstract}

We propose a regular classical field theory realisation of the
Dvali--Gabadadze--Porrati mechanism by considering our universe to be
the four-dimensional core of a seven dimensional 't~Hooft--Polyakov
hypermonopole. We show the existence of metastable gravitons trapped
in the core. Their mass spectrum is discrete, positive definite, and
computed for various values of the field coupling constants: the
resulting Newton gravity law is seven-dimensional at small and large
distances but can be made four-dimensional on intermediate length
scales. There is no need of a cosmological constant in the bulk, the
spacetime is asymptotically flat and of infinite volume in the
extra-dimensions. Confinement is achieved through the local positive
curvature of the extra-dimensions induced by the monopole-forming
fields and for natural values of the coupling constants of order
unity.

\end{abstract}
\maketitle

\section{Introduction}
\label{sec:intro}

Gravity occupies a central role in high energy physics and
cosmology. On one hand, the unification of the fundamental
interactions in the context of String Theory suggests that we may live
in a more than four-dimensional world~\cite{Nordstrom:1914,
  Polchinski:1998rq}. On the other hand, the recent acceleration of
our universe has been confirmed by different experiments and it is now
a widely accepted important result of modern observational
cosmology~\cite{Komatsu:2008hk}. The idea that such an unexplained
acceleration may be the signature of extra-dimensions has been
intensively explored in the recent years~\cite{Arkani-Hamed:1998rs,
  Randall:1999vf, Deffayet:2001pu}. In the Dvali--Gabadadze--Porrati
(DGP) model, the extra-dimensions (bulk) may actually be non-compact
and of infinite volume~\cite{Dvali:2000hr, Dvali:2000xg}. Gravitons
are reflected back onto our universe (brane) due to a different
gravity coupling constant on the brane and in the bulk. The original
DGP action in $\nextra + 4$ dimensions reads
\begin{equation}
\label{eq:action_dgp}
S=\dfrac{\Mp^2}{2} \int{|\sqrt{\gbrane}|} \Rbrane \ud^4 x +
\dfrac{\Mpbulk^{2+\nextra}}{2} \int{ \sqrt{|g|} R \ud^{\nextra+4} X},
\end{equation}
where $\gbrane$ and $\Rbrane$ are respectively the determinant and
scalar curvature of the induced metric along our brane, while $g$ and
$R$ are the corresponding quantities in the bulk. It has been shown
that this model could actually explain the observed acceleration of
the universe, although some works suggest that it may be spoiled by
instabilities~\cite{Deffayet:2000uy, Deffayet:2006wp,
  Gregory:2007xy}. Although the form of Eq.~(\ref{eq:action_dgp}) has
been originally explained by quantum effects, ``regularised models''
have been proposed to justify it from a more classical and tractable
point of view, free of instabilities. Such an approach has been
explored in Refs~\cite{Kolanovic:2003am, Kolanovic:2003da,
  Shaposhnikov:2004ds} and shown to confine gravitons by explicitly
choosing some profile for $\Mpbulk(X)$ or $g(X)$. In a complete
physical framework, both of these functions are however not free and
it is not clear that the DGP mechanism could indeed appear in any
classical system. This question is of crucial importance in order to
assess the viability of both infinite volume extra-dimensions and
instability-free DGP-like mechanism.

In this letter, we answer this question in the context of canonical
classical field theory. Our approach is motivated by condensed matter
physics: topological defects are a direct consequence of the symmetry
breaking mechanism and can model smooth branes~\cite{Kibble:1976,
  Akama:1982jy, Rubakov:1983bb}. Assuming the space-time to be
seven-dimensional, an $SO(3)$ spontaneous symmetry breaking in
$\nextra=3$ codimensions generically forms 't~Hooft--Polyakov
hypermonopoles~\cite{'tHooft:1974qc,Polyakov:74}. In the following, we
prove the existence of a DGP-like mechanism in the core (assumed to be
our universe) of such a monopole.

Compared to lower dimensional defects~\cite{Ringeval:2004ju}, the
existence of positively curved $\nextra-1$ dimensional regions in the
bulk is crucial to allow metastable gravitons to be trapped inside the
core. Six is indeed the minimal number of spatial dimensions for which
there exists a foliation of the extra-dimensions by two-dimensional
positively curved surfaces. In order to allow for a varying Planck
mass, we have for completeness included a dilaton $\dilaton$ having a
mass $\md$ in the Einstein frame. In the Jordan frame, the action
associated with this system is
\begin{equation}
\label{eq:action_pole}
\begin{aligned}
  S & = \dfrac{1}{2 \kappa^2} \int \ue^{\dilaton}\sqrt{-g}  \left[R -g^{AB}
    \partial_A \dilaton \partial_B \dilaton - \dilapot(\dilaton)
    \right] \ud^7 x \\ & + \int \sqrt{-g} \left[-\dfrac{1}{2} g^{AB} \calD_A
    \higgsvect \cdot \calD_{B} \higgsvect -\dfrac{1}{4} \vect{H}_{AB}
    \cdot \vect{H}^{AB} \right. \\ & - \left. \dfrac{\lambda}{8}
    \left(\higgsvect \cdot \higgsvect - \vev^2 \right)^2
    \right]\ud^7 x ,
\end{aligned}
\end{equation}
where the dilaton potential reads $\dilapot=\md^2 \dilaton^2
\exp(2\dilaton/5)$. The $SO(3)$ Higgs field $\higgsvect=\{\higgs^a\}$
is in the triplet representation ($a \in \{1,2,3 \}$). Its vacuum
expectation value $\vev$ breaks $SO(3)$ into $U(1)$. The covariant
derivatives $\calD_A$ enforce gauge invariance and incorporate the
gauge fields $\vect{C}_A=\{C_A^a\}$,
\begin{equation}
  \calD_A \higgsvect  = \partial_A \higgsvect - q \vect{\gauge}_A
  \wedge \higgsvect,
\end{equation}
$q$ being the charge, while the field strength tensor $\vect{H}_{AB}$
is
\begin{equation}
  \vect{H}_{AB}  = \partial_A \vect{\gauge}_B - \partial_B
  \vect{\gauge}_A - q \vect{\gauge}_A \wedge \vect{\gauge}_B.
\end{equation}
As for the dimensional analysis, we have $[\kappa^2]=M^{-5}$,
$[q]=M^{-3/2}$, $[\lambda]=M^{-3}$ and $[C_A]=[\Phi]=[v] = M^{5/2}$.

\section{Background geometry}

\begin{figure}
\begin{center}
\includegraphics[width=0.43\textwidth]{polefields.eps}
\includegraphics[width=0.5\textwidth]{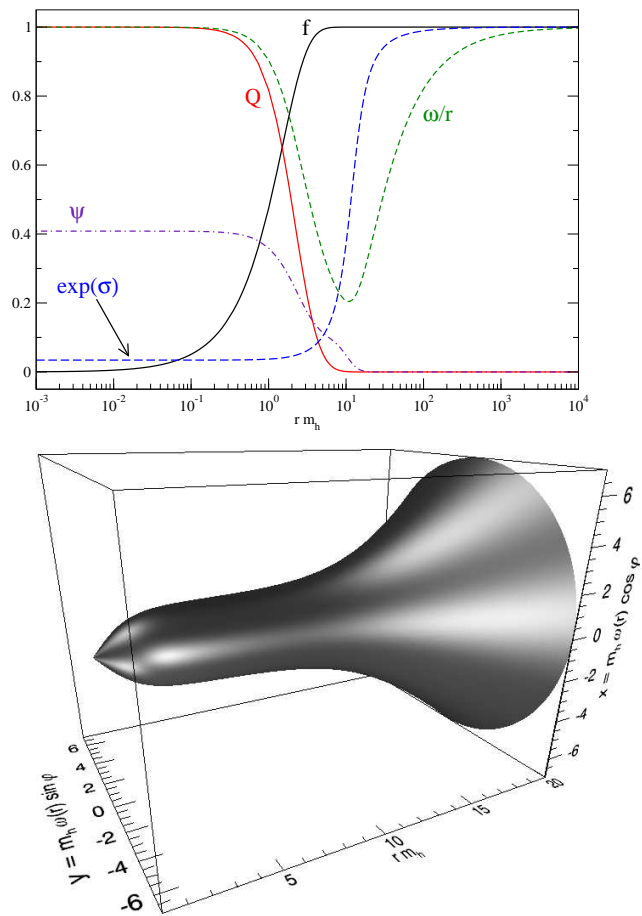}
\caption{Field and metric profiles forming the hypermonopole for
  $\alpha=2.05$, $\epsilon=0.50$ and $\beta=1.00$ (top). The space-time is
  flat asymptotically and in the core, but strongly curved in the
  intermediate region. The spatial sections for $\theta=\pi/2$ are
  represented in the bottom panel as a function of the radial
  coordinate.}
\label{fig:omega}
\end{center}
\end{figure}

Static self-gravitating monopole configurations associated with the
action (\ref{eq:action_pole}) can be obtained by imposing isotropy in
the extra-dimensions, plus Poincar\'e invariance along the four
internal brane coordinates $x^\mu$. Our ansatz for the metric is
\begin{equation}
\label{eq:metric}
\ud s^2 = \ue^{\sigma(r)} \eta_{\mu \nu} \ud x^\mu \ud x^\nu + \ud r^2 +
\omega(r)^2 \ud \Omega^2,
\end{equation}
where $r,\theta,\varphi$ are spherical coordinates in the three
extra-dimensions and $\ud \Omega^2 = \ud \theta^2 + \sin^2 \theta \ud
\varphi^2$. For the monopole-forming fields, the internal space of
$SO(3)$ is mapped to the three extra-dimensions with a purely radial
Higgs field
\begin{equation}
\higgsvect = \vev f(r) \vect{u}_r,
\end{equation}
and winding gauge fields
\begin{equation}
\vect{C}_\theta = \dfrac{1-Q(r)}{q} \vect{u}_\varphi, \quad
\vect{C}_\varphi = - \dfrac{1-Q(r)}{q} \sin \theta  \vect{u}_\theta,
\end{equation}
the other components vanishing. Here, $f(r)$ and $Q(r)$ are two
dimensionless functions such that, far from the core, $f(r)
\rightarrow 1$ and $Q(r) \rightarrow 0$ to recover a Dirac
monopole. In the core, regularity imposes $f(0)=0$ and
$Q(0)=1$. Concerning the metric coefficients, the energy associated
with the defect being finite and localised, we look for asymptotically
flat spacetime, $\sigma \rightarrow 0$, $\omega \rightarrow r$ and
$\dilaton \rightarrow 0$. Regularity in the core also imposes
$\sigma'(0)=\dilaton'(0)=0$ and $\omega \sim r$.

The system of coupled non-linear differential equations obtained from
the action (\ref{eq:action_pole}) is of order ten and does not have
any obvious analytical solution. Once the radial coordinate is
expressed in unit of the Higgs Compton wavelength, the differential
system is parametrised by three dimensionless parameters
\begin{equation}
  \alpha \equiv \kappa^2 \vev^2, \quad \epsilon \equiv
  \dfrac{q^2\vev^2}{\lambda
    \vev^2} = \dfrac{\mb^2}{\mh^2}, \qquad \beta \equiv
  \dfrac{\md^2}{\lambda \vev^2} = \dfrac{\md^2}{\mh^2},
\end{equation}
where $\mh$ and $\mb$ are respectively the mass of the Higgs and gauge
bosons. Under the above-mentioned boundary conditions, the numerical
integration of the equations of motion is a challenging problem that
has been overcome by using recent advances in the
field~\cite{Cash:2005aa}. We have found monopole solutions for almost
any values of the above parameters; only when the stress energy
becomes super-Planckian the system develops some singularities
preventing the spacetime to be asymptotically flat. As can be seen in
Fig.~\ref{fig:omega}, the Higgs and gauge field profiles are typical
of topological defect configurations while the dilaton is
gravitationally trapped inside the core. The profile of $\sigma(r)$
traces the gravitational redshift: clocks are ticking differently
inside and outside the monopole.  More interesting is the profile of
$\omega(r)$. Up to a $4 \pi$ factor, $\omega^2(r)$ gives the area of
the two-sphere of radius $r$ in the extra-dimensions. As can be seen
in Fig.~\ref{fig:omega}, there is a region at finite distance from the
core where $\omega(r)$ does no longer grow as $r$ but remains almost
stationary: the extra-dimensions become cylindrically shaped. As we
show in the next section, gravitons become resonant at these length
scales and metastable from a four-dimensional point of view. Notice
that the spacetime is non-compact and asymptotically Minkowski.

\section{Tensor fluctuations}

We now consider the four-dimensional tensor perturbations around the
previously computed background. The perturbed metric is given by
Eq.~(\ref{eq:metric}) upon the replacement $\eta_{\mu \nu} \rightarrow
\eta_{\mu\nu} + h_{\mu\nu}$, where $h_{\mu\nu}$ is a spacetime
dependent transverse and traceless tensor. The linearised equations of
motion for $h_{\mu\nu}$ are obtained by expanding
Eq.~(\ref{eq:action_pole}) at second order and have already been
derived for an arbitrary number of extra-dimensions in
Ref.~\cite{Ringeval:2004ju}. Defining the dimensionless conformal
radius $z(r)$ and tensor $\xi_{\mu\nu}$ as
\begin{equation}
\label{eq:xidef}
z \equiv \mh \int \exp(-\sigma/2) \ud r, \quad \xi_{\mu \nu} \equiv
\ue^{\dilaton/2} \ue^{3 \sigma/4} \omega h_{\mu\nu},
\end{equation}
the equation of motion for the spin-two fluctuations can be recast
into
\begin{equation}
\label{eq:ximotion}
  -\dfrac{\ud^2 \xi}{\ud z^2} + \left(W^2 + W' -
    \dfrac{\ue^\sigma}{\mh^2 \omega^2} L^2 -\Box \right) \xi = 0, 
\end{equation}
where the tensor indices have been omitted. Derivatives are with
respect to $z$, $\Box=\mh^{-2}\eta^{\mu\nu}\partial_\mu \partial_\nu$
is the d'Alembertian along the brane, and
\begin{equation}
  W = \dfrac{3}{4} \sigma' + \dfrac{\omega'}{\omega} + \dfrac{1}{2}
  \dilaton',\quad L^2 = \partial_\theta^2 +
  \dfrac{\partial_\theta}{\tan \theta} + \dfrac{\partial_\varphi^2}{\sin^2\theta}\,. 
\end{equation}
After a four-dimensional Fourier transform on the brane coordinates,
and an expansion over the spherical harmonics in the angular
extra-dimensions, we have in unit of the Higgs mass $\Box \rightarrow
-\eta_{\mu\nu}p^\mu p^\nu=M^2$ and $L^2 \rightarrow
-\ell(\ell+1)$. One immediately recognises in Eq.~(\ref{eq:ximotion})
the Schr\"odinger equation of a supersymmetric quantum mechanical
system in a central potential $V_2=W^2+W'$~\cite{Cooper:1994eh}. The
operator $L^2$ is the angular momentum, $W(z)$ is the superpotential
and $M^2$ plays the r\^ole of the energy. A subtlety is that our
coordinate $z$ lies on the positive axis only. However, since $h_{\mu
  \nu}$ must remain finite on the brane, Eq.~(\ref{eq:xidef}) implies
that $\xi_{\mu\nu}$ should vanish in $z=0$. Under this condition, and
the usual normalisability at infinity, the differential operators
remain regular enough to use the results of supersymmetric quantum
mechanics. For $M^2=L^2=0$, Eq.~(\ref{eq:ximotion}) is solved by the
``ground state'' $\xi_0 \propto \omega \exp(3 \sigma/4 + \dilaton/2)$
which is however not normalisable asymptotically. The ground state of
the superpartner potential $V_1 = W^2-W'$ is $1/\xi_0$, which is not
regular in $z=0$. As a result, ``supersymmetry'' is broken and the
spectrum is necessarily positive definite, $M^2 > 0$: there is no
massless mode neither tachyon on the brane.

\begin{figure}
\begin{center}
\includegraphics[width=0.43\textwidth]{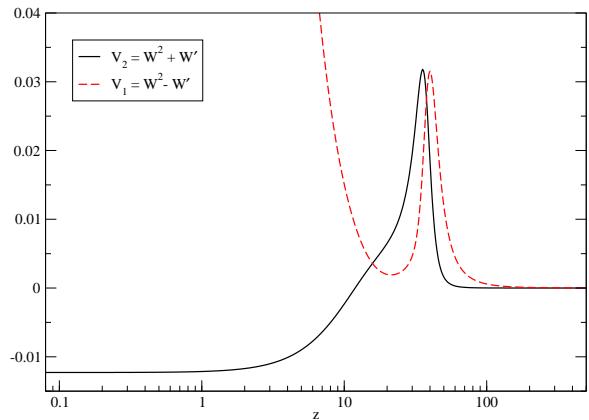}
\caption{Superpartner potentials $V_2(z)$ and $V_1(z)$ confining the
  metastable gravitons. The background fields are those of
  Fig.~\ref{fig:omega}.}
\label{fig:pots}
\end{center}
\end{figure}

Introducing the orthonormal basis of eigenfunctions
$\umode{M}{\ell}(z)$, solutions of Eq.~(\ref{eq:ximotion}), such that
\begin{equation}
  \int_0^\infty \umode{M}{\ell}^*(z_1) \umode{M}{\ell}(z_2)\ud M
= \delta(z_1 - z_2),
\end{equation}
one can check that the retarded Green function for $\xi$ can be
expanded as~\cite{Garriga:1999yh}
\begin{equation}
\label{eq:greenxi}
\begin{aligned}
\green{X_1;X_2} & = - \int \dfrac{\ud^4 p}{(2 \pi)^4} \ue^{i
  p_\mu(x_1^\mu - x_2^\mu)} \sum_{\ell,m}
\Ylm{\ell}{m}(\theta_1,\varphi_1) \\
& \times \Ylm{\ell}{m}^*(\theta_2,\varphi_2) \int
\dfrac{\umode{M}{\ell}(z_1) \umode{M}{\ell}^*(z_2) \ud M}{M^2 +
  (\vec{p})^2 - \left(p^0 + i \epsilon\right)^2}\,.
\end{aligned}
\end{equation}
Using the above equation together with Eq.~(\ref{eq:xidef}), the
tensor modes sourced by any transverse and traceless stress tensor
$S_{\mu \nu}(X)$ are given by
\begin{equation}
\label{eq:hmunu}
\begin{aligned}
&  h_{\mu\nu}(X_1) =-\dfrac{2 \kappa^2}{\mh^2\omega(z_1)}
  \ue^{-\dilaton(z_1)/2}\ue^{-3\sigma(z_1)/4} \\ & \times \int \green{X_1;X_2}
  \ue^{-\dilaton(z_2)/2} \ue^{ + 3 \sigma(z_2)/4} \omega(z_2)
  S_{\mu\nu}(X_2) \ud^7 X_2. 
\end{aligned}
\end{equation}

In order to gain some intuition on the previous expressions, let us
first consider the case of a seven-dimensional flat spacetime. Setting
$\dilaton=\sigma=0$ everywhere, as well as $\omega=r$,
Eq.~(\ref{eq:ximotion}) can be integrated and the normalised modes are
\begin{equation}
\umodeflat{M}{\ell}(z) = \sqrt{M z} \, \besselJ{\ell + 1/2}(M z).
\end{equation}
Considering a ``point-like'' static source $s_{\mu\nu}(x)$ on the
brane
\begin{equation}
S_{\mu \nu}(X) =
\lim_{z\rightarrow 0} \dfrac{1}{z^2} \delta(z) \delta(\cos \theta) \delta(\varphi)
s_{\mu\nu}(x),
\end{equation}
we can explicitly integrate Eqs.~(\ref{eq:greenxi}) and
(\ref{eq:hmunu}) using the flat modes $\umodeflat{M}{\ell}$. Only the
s-waves ($\ell=0$) have a non-vanishing contribution on the brane
since for $z\rightarrow 0$, one has $\umodeflat{M}{\ell\ne0}(z)/z
\rightarrow 0$. The four-dimensional integral over the momentum $p$ in
Eq.~(\ref{eq:greenxi}), together with the denominator containing
$M^2$, is the classical retarded Yukawa propagator. After some
calculations, one finally gets on the brane
\begin{equation}
\label{eq:hmunuflat}
\begin{aligned}
  h_{\mu\nu}^{\flat}(\vec{x}_1) & = \lim_{z\rightarrow 0}
  \dfrac{\kappa^2}{8 \pi^2 \mh^2} \int \ud^3 \vec{x}_2 \,
  s_{\mu\nu}(\vec{x}_2)  \\
  & \times \int \ud M \dfrac{\left|\umodeflat{M}{0}(z)\right|^2 }{z^2}
  \dfrac{\ue^{-M \left|\Delta\vec{x}\right|}}{\left|\Delta
      \vec{x}\right|}\, ,
\end{aligned}
\end{equation}
where $\Delta \vec{x} \equiv \vec{x}_1 - \vec{x}_2$. Using the
expansion of the Bessel function, $\umodeflat{M}{0}(z) \sim
\sqrt{2/\pi} M z$, the previous expression simplifies to
\begin{equation}
h_{\mu \nu}^{\flat} = \dfrac{2\kappa^2}{4 \pi^3\mh^2} \int \ud^3 \vec{x}_2
\dfrac{s_{\mu\nu}(\vec{x}_2)}{\left| \Delta \vec{x} \right|^4}\,,
\end{equation}
which is the standard linearised solution of the Einstein equations
around a seven-dimensional Minkowski spacetime. Notice the power law
dependence $1/|\Delta \vec{x}|^{d-2}$ in $d=6$ spatial dimensions, as
well as the $4\pi^3$ factor which is $d-2$ times the surface of the
unit $(d-1)$--sphere, as one would have obtained from the Gauss law.

\begin{figure}
\begin{center}
\includegraphics[width=0.43\textwidth]{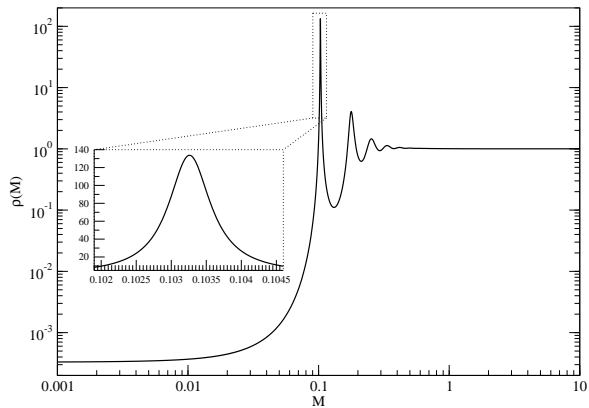}
\caption{Spectral density $\rho(M)$ as a function of the graviton mass
  $M$. There are at least two trapped metastable gravitons in the
  background geometry of Fig.~\ref{fig:omega}. The long lived resonance
  is well fitted by a Breit--Wigner distribution centered at
  $\mg\simeq 0.10326$ with a width $\Gamma \simeq 3.5\times10^{-4}$ (in
  $\mh$ units).}
\label{fig:spectrum}
\end{center}
\end{figure}

In the background geometry of the hypermonopole, the situation is
nearly the same apart that the mode functions $\umode{M}{\ell}(z)$
are now modified. As can be seen in Eq.~(\ref{eq:hmunuflat}), we need
the values of the rescaled spectral density associated with s-waves on
the brane $\rho(M)=|\umode{M}{0}(0)|^2/|\umodeflat{M}{0}(0)|^2$. The
tensor modes are then given by
\begin{equation}
\label{eq:hmunupole}
  h_{\mu \nu} = \dfrac{2\kappa^2 \ue^{-\dilaton(0)}}{8 \pi^3 \mh^2} \int \ud^3
  \vec{x}_2  \dfrac{\laplace{\rho(M)M^2}} {\left| \Delta \vec{x} \right|}
  s_{\mu\nu}(\vec{x}_2),
\end{equation}
where $\laplace{.}$ stands for the forward Laplace transform evaluated
at $|\Delta\vec{x}|$. From the monopole-forming fields computed in the
previous section, we have plotted in Fig.~\ref{fig:pots} the potential
$V_2(z)$ and its superpartner. Since the spacetime is asymptotically
flat, $V_2$ vanishes at infinity and there is not any bound
state. However, $V_2$ (and also $V_1$) exhibits a barrier at the
location of maximum curvature allowing metastable modes in the
core. We have numerically solved the equation of motion
(\ref{eq:ximotion}) for these potentials and plotted in
Fig.~\ref{fig:spectrum} the resulting spectral density. When this
quantity is constant, gravity is purely seven-dimensional on the
brane. This is the case for large, but also for low values of $M^2$,
as in the Gregory--Rubakov--Sibiryakov
model~\cite{Gregory:2000jc}.

Notice that the constant value of $\rho$ gives the effective
gravitational coupling constant: here, it is different for large and
low values of $M$ due to the dilaton condensation in the core as well
as the gravitational redshift. In the intermediate range, $\rho(M)$ is
strongly peaked for particular values of $M$: these are the resonant
metastable modes. We have also numerically checked that there is not
any bound state with $M^2\le 0$, as expected from the supersymmetry
arguments. Changing the background parameters $\alpha$, $\beta$ and
$\epsilon$ affects the mass spectrum and the width or number of
metastable modes can be easily adjusted. Lowering $\epsilon$
delocalises the gauge fields (in unit of the Higgs Compton wavelength)
and the position of the barrier is pushed towards larger values of
$z$. Decreasing $\alpha$, or increasing $\epsilon$, reduces the height
of the barrier while $\beta$ changes its shape. Let us notice that if
$\alpha$ is too small, or $\epsilon$ too big, we do no longer observe
resonances and this corresponds to the disappearance of the confining
nest on $V_2$. This is reminiscent with the properties of bound states
in the case of broken supersymmetry.

To understand how these resonances realise the DGP mechanism, one can
approximate them as Dirac distributions. Let us say we have one
trapped graviton at a mass $\mg$, then neglecting the smooth changes
in the spectral density, $\rho(M)\simeq 1+ \const \delta(M-\mg)$ where
$\const$ encodes how peaked the resonance is. The Laplace transform in
Eq.~(\ref{eq:hmunupole}) simplifies to
\begin{equation}
  \laplace{M^2 \rho(M)} = \dfrac{2}{|\Delta \vec{x}|^3}
  + \const \mg^2 \ue^{-\mg
    |\Delta \vec{x}|},
\end{equation}
and four-dimensional gravity is recovered over the length scales
$(\mg/\const)^{1/3} <|\Delta \vec{x}| \mg < 1$, provided the mode is
light enough $\mg < \const$. At small and large distances, seven
dimensional gravity is recovered while in between we have even
observed some fractional power dependencies. When more than one
gravitons are trapped the situation becomes even more complex and the
detailed analysis of these effects is left for a forthcoming work.

\section{Conclusion}

We proposed here a canonical classical field theory model which
describes a seven-dimensional monopole, at the core of which gravitons
get trapped. Their mass spectrum being positive definite, there are no
instabilities for the tensor modes. This phenomenon turns out to be a
natural way, in the context of field theory, to implement the DGP
idea. The required field configurations can be obtained without
fine-tuning from a dense set of coupling constant values of order
unity. However, to obtain a four-dimensional gravity behaviour over a
wide range of length scales, some amount of fine-tuning is certainly
required to confine an almost massless mode. Notice that the smaller
the graviton mass, the larger the effective four-dimensional Planck
mass, possibly addressing the mass hierarchy problem. It would be
interesting, if possible, to find a condensed matter system for which
this mechanism could be experimentally explored.

\begin{acknowledgments}

This work is supported by the Belgian Federal Office for Science,
Technical and Cultural Affairs, under the Inter-university Attraction
Pole grant P6/11.

\end{acknowledgments}

\bibliography{bibpole}

\end{document}